\documentclass[runningheads]{llncs}
\usepackage[T1]{fontenc}
\usepackage{algorithmic}
\usepackage{algorithm}
\usepackage{amsmath,amssymb,amsfonts}
\usepackage{bm}
\usepackage{booktabs,makecell,multirow,array,colortbl,dcolumn,stfloats}
\usepackage[font=small]{caption}
\usepackage{float}
\usepackage{graphicx}
\graphicspath{{figs/}}
\usepackage{lscape}
\usepackage{multirow}
\usepackage{subcaption}
\usepackage{textcomp}

\usepackage{tabularx}

\title{Agent-Based Simulation of a Financial Market with Large Language Models}

\author{
    Ryuji Hashimoto\inst{1}\orcidID{0009-0008-0042-1477} \and
    Takehiro Takayanagi\inst{1}\orcidID{0009-0000-6467-8222} \and
    Masahiro Suzuki\inst{1}\orcidID{0000-0001-8519-5617} \and
    Kiyoshi Izumi\inst{1}\orcidID{0000-0003-0870-7310}
}
\authorrunning{R. Hashimoto et al.}
\institute{Simulacra Inc. \and The University of Tokyo
\email{hashimoto-ryuji419@g.ecc.u-tokyo.ac.jp}}

\begin{document}
\maketitle
\begin{abstract}

In real-world stock markets, certain chart patterns---such as price declines near historical highs---cannot be fully explained by fundamentals alone. These phenomena suggest the presence of path dependence in price formation, where investor decisions are influenced not only by current market conditions but also by the trajectory of prices leading up to the present. Path dependence has drawn attention in behavioral finance as a key mechanism behind such anomalies.
One plausible driver of path dependence is human loss aversion, anchored to individual reference points like purchase prices or past peaks, which vary with personal context. However, capturing such subtle behavioral tendencies in traditional agent-based market simulations has remained a challenge.
We propose the Fundamental-Chartist-LLM-Agent (FCLAgent), which uses large language models (LLMs) to emulate human-like trading decisions. In this framework, (1) buy/sell decisions are made by LLMs based on individual situations, while (2) order price and volume follow standard rule-based methods.
Simulations show that FCLAgents reproduce path-dependent patterns that conventional agents fail to capture. Furthermore, an analysis of FCLAgents' behavior reveals that the reference points guiding loss aversion vary with market trajectories, highlighting the potential of LLM-based agents to model nuanced investor behavior.

\keywords{LLMs \and Financial market simulations \and Behavioral biases}
\end{abstract}

\section{Introduction}
An agent-based market simulation is an effective tool for modeling macro-scale financial phenomena. Researchers aim to constructively gain insight regarding the underlying mechanisms of these phenomena by designing investor behavior using heterogeneous agents and simulating their interactions~\cite{am3,am1,am4,am6}. 

One of the key factors in modeling investor behavior as agents is the incorporation of behavioral biases. Agent-based approach focuses on bounded rationality~\cite{bounded_rationality} to capture dynamics of complex systems such as financial markets. Unlike neoclassical economics, which assumes fully rational agents who optimize utility, the agent-based perspective requires behavioral principles that are not necessarily based on optimization are required. Behavioral biases represent one manifestation of human behavior that reflects such bounded rationality. Research in behavioral finance~\cite{behavioral_biases_survey1,behavioral_biases_survey2,behavioral_biases_survey3} has identified various behavioral biases among investors, suggesting that market participants' irrationality follows certain systematic patterns. 
Several studies in agent-based modeling have examined the impact of incorporating behavioral biases, especially focusing on loss aversion~\cite{pt_inspired_agent1,pt_inspired_agent2,pt_inspired_agent3}, into agent models to assess their effects on macro-level phenomena. Loss aversion refers to the tendency of individuals to weigh potential losses more heavily than equivalent gains, often leading to risk-averse or -seeking behavior depending on the framing of outcomes.

Despite their utility, market simulation approaches face a significant challenge in incorporating context-dependent behavioral biases into agents, limiting their ability to fully capture real-world financial dynamics. In financial markets, investor behavior depends not only on current situations---such as current market price or individual's holding positions, but also on context. For instance, the reference point that distinguishes perceived gains from losses---a crucial hyperparameter in formalizing loss aversion---is not uniquely defined and varies based on trading and market history~\cite{reference_point_difficulty}. Although this tendency is believed to help explaining market phenomena that deviate from neoclassical assumptions---such as the path dependence of asset prices---traditional agent-based models struggle to capture the complex nuances of behavioral biases, as they rely on predefined and static representations of decision rules.

Given this challenge, a promising alternative is the use of large language models (LLMs). In recent years, LLMs have emerged as advanced artificial intelligence systems trained on vast amounts of text data to understand and generate natural language. Since LLMs are trained on diverse human-generated texts, there is growing speculation that they may exhibit behavioral biases similar to those of humans~\cite{hagendorff2023human}. If the context-dependent behavioral biases are embedded in LLMs, LLM-based agent models could represent a breakthrough in multi-agent market simulations, reproducing more human-like behavioral patterns and elucidating connection between such behavior and macro-level market outcomes.

\begin{figure*}[tbp]
\centering
\includegraphics[width=0.95\linewidth]{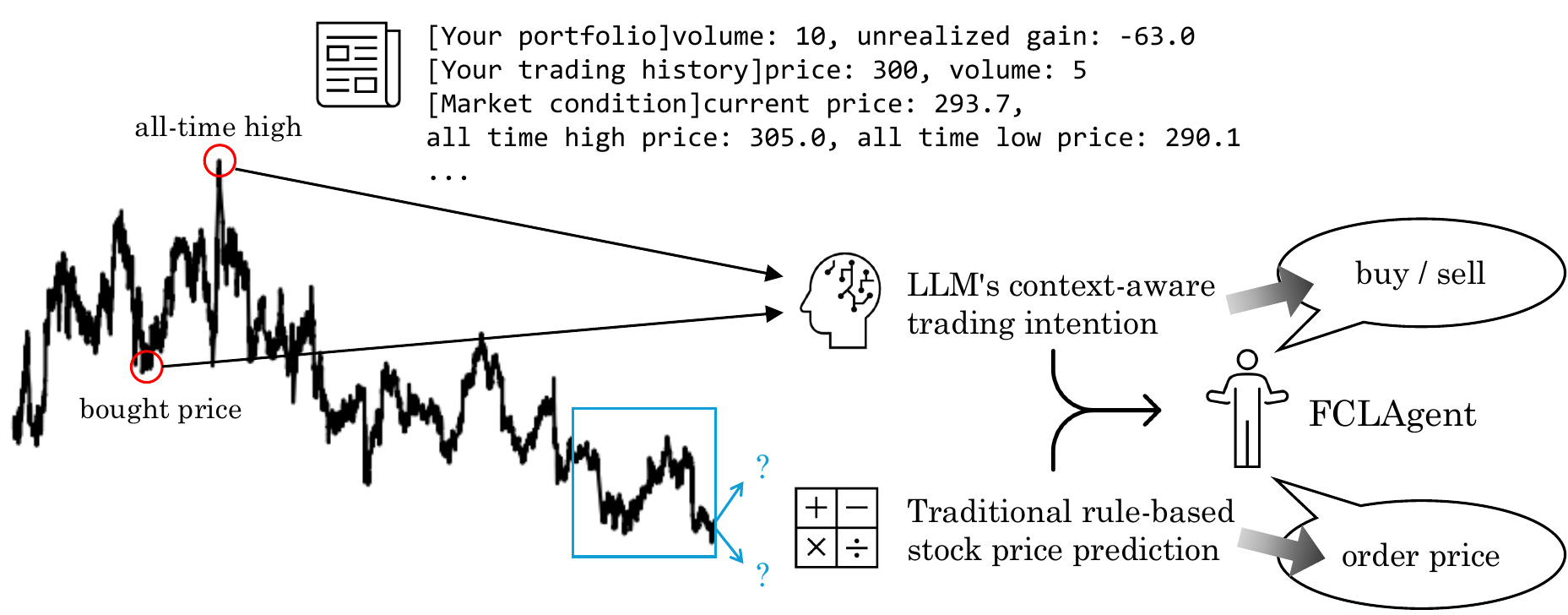}
\caption{Schematic overview of the FCLAgent architecture. The agent integrates a psychologically biased trading intention generated by an LLM with a traditional rule-based mechanism for stock price prediction. This hybrid structure allows the agent to exhibit human-like decision patterns while maintaining numerical reliability.}
\label{Fig:idea_fcl_agent}
\end{figure*}

Although the potential of LLMs in financial market simulations is promising, their application remains limited. This is due to the lack of methods for extracting plausible behavioral tendencies while avoiding well-known limitations in numerical reasoning~\cite{llm_numerical_reasoning_limitation}. This study proposes a novel agent model, Fundamental-Chartist-LLM-Agent (FCLAgent), which integrates context-dependent loss-averse behavior generated by LLMs into trading intention, while relying on traditional rule-based mechanisms for order pricing. The main idea of the FCLAgent is illustrated in Fig.~\ref{Fig:idea_fcl_agent}. The FCLAgent is a new variant of FCNAgent~\cite{fcn_agent}, a standard agent model commonly used in stock market simulations. The FCLAgent decouples the decision-making process into two components: the decision of whether to buy or sell, which is guided by LLM-generated responses to contextual prompts and reflects behavioral tendencies such as loss aversion, and the order execution, which is governed by predefined, rule-based logic inherited from conventional agent model~\cite{fcn_agent}. The trading intention is expressed as a text-formatted order direction (buy or sell), while the actual order price and volume are computed through deterministic rules based on market conditions. This hybrid structure enables FCLAgent to exhibit psychologically plausible behavior derived from the LLM outputs, without relying on the LLM to perform numerical operations.

In our experiments, we evaluated the effectiveness of LLMs in financial market simulations from both macro- and micro-level perspectives. Macro level analysis demonstrated that the inclusion of the FCLAgents enables the emergence of empirically observed market anomalies, such as the negative correlation between proximity to the all-time high and future returns~\cite{reference_point_ath1}. At the micro level, we analyzed the behavioral patterns of the FCLAgents to see whether their trading decisions are affected by the all-time high, while they do not exhibit systematic imbalances in buy/sell frequency or portfolio allocation---features that support their plausibility as investor models. Additionally, we analyzed LLMs' decision-making across four different contexts with varying potential reference points in single-turn simulations. The results revealed that some types of LLMs exhibited human-like behavioral patterns, such as a tendency to realize gains more readily than losses, and to adjust their risk preferences based on multiple reference points including purchase price and all-time high or low.

\section{Related Work}

\subsection{Behavioral Biases in Agent-Based Simulations}
Since the global financial crisis, multi-agent market simulation techniques have attracted attention as constructive approach for investigating financial markets as complex systems~\cite{am1}. A central aspect of these simulations is the agent modeling. To ensure the validity and relevance of simulation outcomes, researchers have developed a wide range of agent models designed to capture typical human behavioral patterns~\cite{odd+d_protocol}, with particular emphasis on behavioral biases such as rational inattention~\cite{rational_inattention}, egocentric bias~\cite{ego_bias_in_mas}, herd behavior~\cite{fcn_agent,herding_into_agent}, and loss aversion~\cite{pt_inspired_agent1,pt_inspired_agent2,pt_inspired_agent3}.

\subsubsection{Loss Aversion in Agent-Based Simulations}\label{Sec:loss_aversion_agent}
Loss aversion~\cite{prospect_theory1,prospect_theory2} describes a systematic asymmetry in human risk preferences: individuals tend to exhibit risk-averse behavior when faced with potential gains, yet become risk-seeking in the context of potential losses, aiming to avert those losses even at greater risk.

In financial markets, loss aversion manifests in investor behavior as a tendency to realize gains prematurely while holding onto losing positions. This phenomenon, referred to as the {\em disposition effect}, has been empirically validated by Odean~\cite{disposition_effect} through the analysis of individual investors' transaction histories. The authors introduced the {\em Proportion of Gains (Losses) Realized} (PGR and PLR, respectively) as measures of realized versus unrealized outcomes, and found that PGR consistently exceeds PLR, indicating a systematic bias in decision-making under gain and loss scenarios.

Despite the well-documented financial implications of loss aversion~\cite{loss_aversion_implication1,loss_aversion_implication2,loss_aversion_implication4,loss_aversion_implication5,loss_aversion_implication3}, incorporating this bias into agent-based models remains a significant challenge. One major difficulty lies in the context-dependent nature of reference points---the benchmarks individuals use to evaluate outcomes as gains or losses~\cite{reference_point_difficulty}. These reference points may vary across individuals and situations, with common examples including the purchase price~\cite{reference_point_bought_price} and the asset's all-time high or low~\cite{reference_point_ath2,reference_point_ath1}.
Consider, for example, an investor who purchases a stock at \$300. If the price subsequently rises to \$400 and then declines to \$350, the perceived outcome depends on the chosen reference point. Using the purchase price, the investor perceives a \$50 gain; using the prior peak of \$400, the same price represents a \$50 loss. The investor’s behavioral response may vary considerably between these two framings, even though the final price is identical. Moreover, recent advancements in social networks are thought to provide information about other investors' purchase prices, offering additional reference points that amplify human investors' {\em fear of missing out} (FOMO)---a form of loss aversion where missed opportunities are perceived as losses~\cite{fomo}. This sensitivity to context complicates efforts to formalize loss aversion within agent models. To date, no existing framework has successfully captured this context-dependency in a systematic and generalizable way.

\subsection{LLMs for Social and Financial Simulations}
LLMs have recently emerged as promising tools for constructing sophisticated agent models in social simulations~\cite{llm_social_science}, owing to their capacity to represent distinct agent profiles, generate text-based behavioral data, and engage in complex multi-agent interactions~\cite{llm_agent_simulation1,llm_agent_simulation2,llm_agent_simulation3}. For instance, micro-societies composed of LLM-based agents were developed and the emergence of sociability among them was examined~\cite{llm_micro_society2,llm_micro_society1}. In addition, LLMs have been employed in simulations of macroeconomic dynamics~\cite{econagent,llm_laboratory_market}, social norms formation~\cite{llm_social_norms}, negotiation processes~\cite{llm_negotiation,llm_negotiation2}, and social network evolution~\cite{llm_sns1,llm_sns2}.
Despite these advancements, the integration of LLMs into agent-based financial market simulations remains underexplored. This is primarily due to two unresolved challenges: (1) the behavioral fidelity required for financial decision-making---such as context-dependent risk sensitivity and bounded rationality---has not been systematically identified or incorporated into LLM-based agents; and (2) there is no established methodology for constructing agent models that both mitigate LLMs' known limitations in numerical reasoning~\cite{llm_numerical_reasoning_limitation} and preserve interpretability.

\section{Proposed Method: FCLAgent}
We propose a novel agent model that distills and refines the human-like behavioral characteristics captured by LLMs, incorporating context-dependent loss aversion into trading intention while delegating order price determination to traditional rule-based mechanisms. This design enables realistic simulations of behavioral biases while circumventing the well-known numerical reasoning limitations of LLMs~\cite{llm_numerical_reasoning_limitation}. In this section, we first desxcribe the structure of our simulation framework and then introduce the agent models employed: the FCNAgent~\cite{fcn_agent,fcn_agent2} and the proposed FCLAgent, which incorporates LLM-derived trading intentions into the order decision rules based on the FCNAgent.

\begin{figure*}[tbp]
\centering
\includegraphics[width=0.95\linewidth]{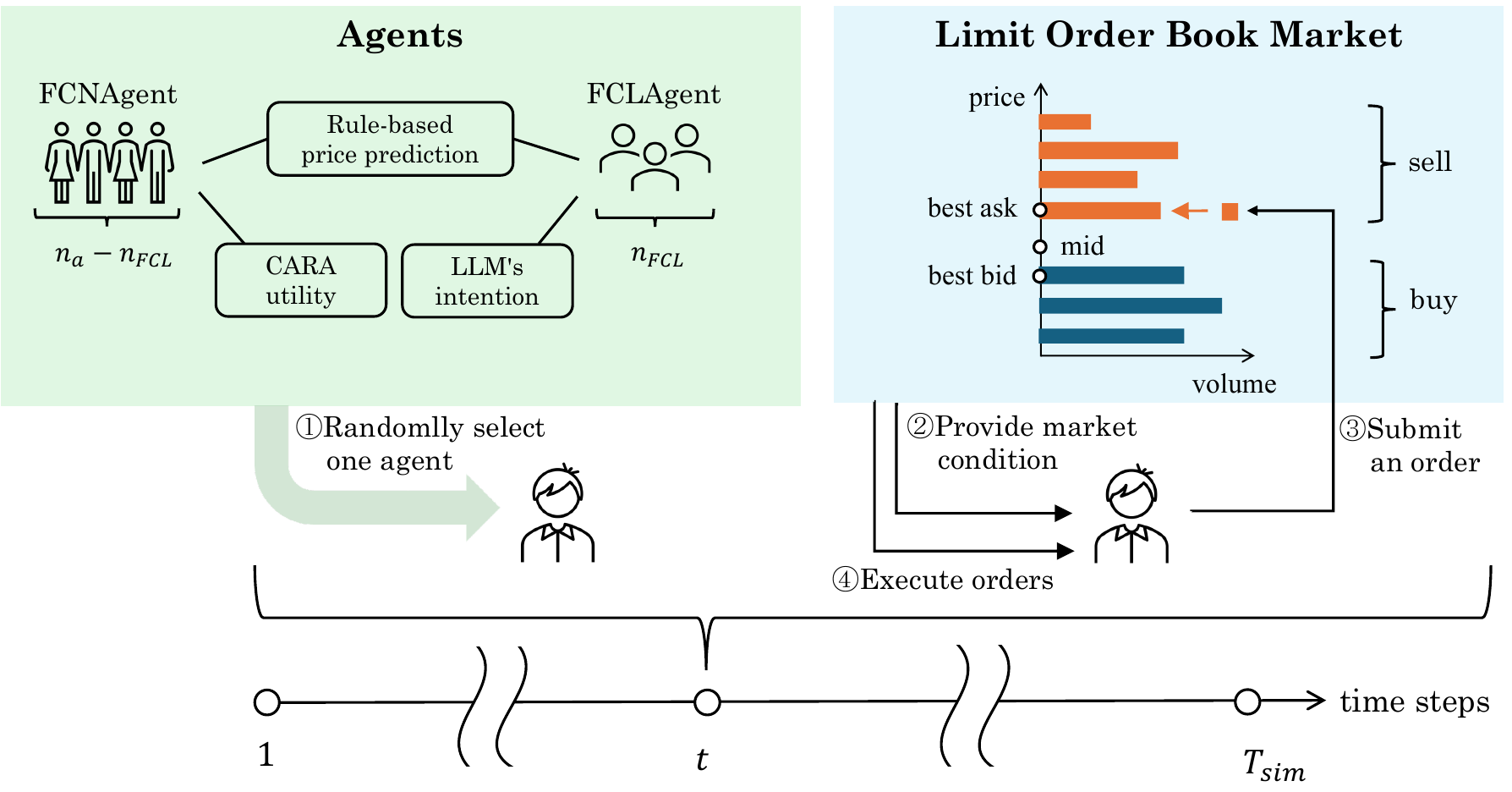}
\caption{Overview of the simulation framework. At each time step $t\in\{1,...,T_{sim}\}$, one of the $n_a$ agent---either an FCNAgent or an LLM-guided FCLAgent---is selected to submit a limit order. Their actions are placed into a limit order book market, which matches buy and sell orders to determine transactions and update the market state.}
\label{Fig:simulation_structure}
\end{figure*}

\subsection{Simulation Structure}
In this work, we assume $n_a\in\mathbb{N}$ agents operate in a single market setting, with each simulation consists of $T_{sim}\in\mathbb{N}$ time steps. Fig.~\ref{Fig:simulation_structure} illustrates the structure of our simulation settings. At each time step $t\in\{1,...,T_{sim}\}$, a randomly selected agent $j\in\{1,...,n_a\}$ is allowed to place an order. Agent $j$ submits an order specifying a signed order volume $v_t^j \in \mathbb{Z}$ and an order price $p_t^j \in \mathbb{R}_+$. The sign of $v_t^j$ indicates whether the agent intends to buy or sell\footnote{For instance, if $v_t^j = -2$ and $p_t^j = 300.0$, agent $j$ submits an order to sell $2$ units of the stock at a price of 300.0.}.

The market operates under the double auction rule, where orders are collected into a limit order book. The limit order book displays anonymized prices and volumes at which agents are willing to buy or sell the stock, and the market matches orders based on price and time priority.

\subsection{Traditional Agent: FCNAgent}
Chiarella and Iori~\cite{fcn_agent} introduced the FCNAgent, a widely adopted agent model for limit order book market simulations. The FCNAgent $j$ predicts future stock returns $\hat{r}_{t+\tau^j}^j$ and price $\hat{p}_{t+\tau^j}^j$ given the current time steps $t$, market price $p_t$, and fundamental price $p_t^f$ as:
{\small
\begin{align}
    \hat{r}_{t+\tau^j}^j &= \frac{1}{w^{j,f} + w^{j,c} + w^{j,n}} \left( \frac{w^{j,f}}{\tau^f} \log \frac{p_t^f}{p_t} 
    + \frac{w^{j,c}}{\tau^j} \log \frac{p_t}{p_{t-\tau^j}} 
    + w^{j,n} \epsilon_t \right) \label{Eq:fcn_return} \\
    \hat{p}_{t+\tau^j}^j &= p_t \exp\left( \tau^j \hat{r}_{t+\tau^j}^j \right) \label{Eq:fcn_price}
\end{align}}
Here, $\tau^j\in\mathbb{N}$ denotes the time window size of agent $j$. $w^{j,f},w^{j,c},w^{j,n}\in\mathbb{R}_+$ are the coefficients corresponding to the three components described below, randomly determined for each agent at the beginning of the simulation such that $w^{j,f}\sim Ex(\lambda^f),~~ w^{j,c}\sim Ex(\lambda^c),~~ w^{j,n}\sim Ex(\lambda^n)$. Here, $Ex(\lambda)$ indicates an exponential distribution with expected value $\lambda\in\mathbb{R}_+$. The first term inside the parentheses in Equation~(\ref{Eq:fcn_return}) $\frac{w^{j,f}}{\tau^f} \log \frac{p_t^f}{p_t}$ is called the fundamental trader component, implying that agent $j$ predicts that market price $p_t$ will reverse to fundamental price $p_t^f$ in $\tau^f$ steps. $\tau^f\in\mathbb{N}$ denotes the mean-reversion time. The second term $\frac{w^{j,c}}{\tau^j} \log \frac{p_t}{p_{t-\tau^j}}$ is called the chartist trader component, modeling a typical ordering pattern, following price trend~\cite{info_effect}. The last term $w^{j,n} \epsilon_t$ is called the noise trader component. $\epsilon_t$ is Gaussian noise with a mean $0$ and variance $(\sigma^n)^2$. FCNAgent was named after its three components, fundamental, chartist, and noise traders, which together form its return prediction model.

Chiarella {\em et al.}~\cite{fcn_agent2} proposed the order decision of the FCNAgent based on price prediction $\hat{p}_{t+\tau^j}^j$, holding cash amount $c_t^j\in\mathbb{R}$, holding stock position $w_t^j\in\mathbb{Z}$, and the risk-aversion term $\alpha^j\in\mathbb{R}_+$ by assuming the constant absolute risk aversion (CARA) utility function $\mathcal{U}^j$:
\begin{align}
\mathcal{U}^j=-\exp\left\{-\alpha^j(w_t^jp_t+c_t^j)\right\}\label{Eq:cara}
\end{align}
where $\forall j~~ c_1^j\sim U(c_{min},c_{max})$,$w_1^j=\lceil w^j\rceil,~w^j\sim U(w_{min},w_{max})$,$c_{min},c_{max},w_{min}$,\\$w_{max}\in\mathbb{R}_+$. $\lceil\cdot\rceil$ denotes the ceiling function and $U(c_{min},c_{max})$ represents the uniform distribution from range $[c_{min},c_{max}]$. $\alpha^j$ and $\tau^j$ are set as:
{\small
\begin{align}
\forall j~~ \alpha^j=\alpha\frac{\alpha^{diff}+w^{j,f}}{\alpha^{diff}+w^{j,c}},~\tau^j=\left\lceil\tau\frac{\tau^{diff}+w^{j,f}}{\tau^{diff}+w^{j,c}}\right\rceil\label{Eq:alpha_tau}
\end{align}}
where $\alpha,\tau\in\mathbb{R}_+$ are the reference levels of $\alpha^j$ and $\tau^j$, and $\alpha^{diff},\tau^{diff}\in\mathbb{R}_+$ are external parameters that control the range of variation in $\alpha^j$ and $\tau^j$.

The agent decides $p_t^j$ and $v_t^j$ to maximize the expected utility $\mathbb{E}_t[\mathcal{U}_{t+\tau^j}^j]$, where $\mathbb{E}_t[\cdot]$ denotes the expected mean given the information available at time $t$.

\subsection{Proposed Agent: FCLAgent}
Instead of CARA derived order price calculation~\cite{fcn_agent2}, the FCLAgent decides whether to buy or sell the stock according to the output of the LLM:
\begin{align}
v_t^j=\mathcal{F}(\iota_t^j)v^j
\end{align}
where $\mathcal{F}:\mathcal{I}\mapsto\{-1,1\}$ is LLM that generate order intention using following information $\iota_t^j\in\mathcal{I}$. $v^j\in\mathbb{N}$ is a constant variable that controlls the order volume of the FCLAgent. $\iota_t^j$ is provided as a text prompt containing the given information. 
\begin{itemize}
\item \textbf{Current portfolio}: Holding cash amount $c_t^j$, holding volume of the stock $w_t^j$, and the unrealized gain $\bar{g}_t^j$. Unrealized gain refers to the increase in value of the stock that has not yet been sold.
\item \textbf{Market condition}: Current stock price $p_t$, All-time high and low prices $p_{1:t}^{h}=\max(p_1,...,p_t),~p_{1:t}^{l}=\min(p_1,...,p_t)$, remaining and total time steps $T_{sim}-t, T_{sim}$, and order flow imbalance $OFI$. $OFI$ is calculated as $\frac{n_{buy}-n_{sell}}{n_{buy}+n_{sell}}$, where $n_{buy}$ and $n_{sell}$ represent the total volume of buy and sell orders displayed in the limit order book at a given time.
\item \textbf{Trading history}: Traded time, price and volume of all passed time $h_t^j=\{t', p_{t'}^j,v_{t'}^j\}_{t'=1}^{t-1}$.
\end{itemize}
An example of the prompt is provided in Appendix~\ref{Appx:prompt}. Unrealized gain of the FCLAgent $j,~ \bar{g}_t^j$ is calculated as the difference between current holding stock value and the total cost paid to get the current position.
\begin{eqnarray}
\bar{g}_t^j= v_{1:t-1}^{total}p_t-\sum_{t'=1}^{t-1}v_{t'}^jp_{t'},~v_{1:t-1}^{total}=\sum_{t'=1}^{t-1}v_{t'}^j\label{Eq:unrealized_gain}
\end{eqnarray}
While utilizing LLM output for order direction, FCLAgent decides their order price based on traditional rule~\cite{fcn_agent}:
\begin{eqnarray}
p_t^j=\begin{cases}
\min\left(\hat{p}_{t+\tau^j}^j(1-m^j), p_t^{ask}\right)~~~ \mathrm{if}~~~0<v_t^j\\
\max\left(\hat{p}_{t+\tau^j}^j(1+m^j), p_t^{bid}\right)~~~ \mathrm{otherwise}
\end{cases}
\end{eqnarray}
where $p_t^{bid},p_t^{ask}$ are the best bid and ask prices. The best bid (ask) price refers to the highest (lowest) price buyers (sellers) are willing to trade on the limit order book. Fig.~\ref{Fig:simulation_structure} describes each price within the limit order book. $m^j$ is a fixed order margin of agent $j$, which is determined as $\forall j~ m^j\sim U(m_{min},m_{max})$.

\section{Experiment}\label{Sec:experiment}
This section presents the experiment conducted to evaluate the effectiveness of the proposed FCLAgent. We perform limit order book market simulations, as illustrated in Fig.~\ref{Fig:simulation_structure}, and test whether simulations incorporating the FCLAgent successfully replicate a well-documented stock market anomaly: the nearness to an asset's all-time high price negatively predicts its future return~\cite{reference_point_ath1}, which was unreplicable by conventional agents .

\subsection{Multi-Agent Simulation}
In our experimental setup, we begin with a baseline simulation composed of $n_a$ FCNAgents. To examine the impact of FCLAgents, we replace $n_{\mathrm{FCL}}$ of the FCNAgents with FCLAgents, where $n_{\mathrm{FCL}}<n_a$. We conduct experiments by varying $n_{\mathrm{FCL}}$ to assess how the proportion of FCLAgents influences market dynamics. We run five trials of simulations with different seed values for every $n_{\mathrm{FCL}}$. The details of the experimental settings are listed as follows.
\begin{itemize}
\item \textbf{Number of time steps}: As shown in Fig.~\ref{Fig:simulation_structure}, a time step is defined as the period during which a randomly selected agent places orders. In our experiment, each simulation consists of $T_{sim}=(100+750+10+750)\times500$ time steps. Here, each $1,610$ time steps is regarded as one trading day, resulting in a total of $500$ simulated days. Within each day, the first $100$ steps and the $10$ steps from step $850$ to $860$ were designated as order collection phases, during which no order matching is performed and only order placement is allowed. We record every order and execution event with several statistics such as market/mid prices and order/execution volumes at each time step to form synthetic tick data\footnote{Tick data refers to high-frequency records of every individual trade or quote, including information such as price, volume, and timestamp. In synthetic tick data, each record is assigned a simulation-based timestamp represented as an integer, whereas in real tick data, timestamps correspond to the actual calendar time at which each event occurred.}.
\item \textbf{Market}: We set the number of markets to $1$, tick size\footnote{Tick size refers to the minimum ordering price unit.} to $1.00\times10^{-2}$, initial market price to $p_0=300.00$, and the fundamental price $p_t^f$ to follow a geometric Brownian motion with zero drift and a volatility of $1.00\times10^{-4}$.
\item \textbf{Agent}: We set the value of following parameters regarding the agent models as follows. $n_a=1,000$, $\lambda^f=10.00$, $\lambda^c=1.50$, $\lambda^n=1.00$, $\sigma^n=1.00\times10^{-2}$, $c_{min}=0.00$, $c_{max}=3.00\times10^4$, $w_{min}=0$, $w_{max}=100$, $\alpha^{diff}=20.00$, $\alpha=0.10$, $\tau^{diff}=30.00$, $\tau=100.00$, and $\tau^f=200.00$. For the FCLAgent\footnote{To amplify the impact of introducing FCLAgents without increasing their number, we assign them a larger trading size per order than the other FCNAgents. This design choice allows us to reduce the number of FCLAgents required in the simulation, thereby shortening the overall computation time while still capturing their behavioral influence on market dynamics.}, $\forall j~ v^j=100$, $c_{min}=0.00$, $c_{max}=1.00\times10^5$, $w_{min}=0$, $w_{max}=300$, $m_{min}=0.00$, and $m_{max}=0.01$. 
\item \textbf{Others}: The study focuses on open-sorce LLM, with Llama 3.1 8B~\cite{llama3} in this experiment. The multi-agent market simulations were implemented primarily using PAMS, a Python-based limit order book market simulator~\cite{pams}.
\end{itemize}
To compare the simulation results with real data, we use FLEX-FULL historical tick data provided by the Japan Exchange Group~\cite{flex_full}. The data period is from January 5, 2015 to August 20, 2021. As described in Table~\ref{Tab:tickers}, we select $18$ stocks from those with sufficiently high liquidity during the period, ensuring a diverse range of industries.

\begin{table*}[tbp]
\centering
\caption{Ticker codes of real data used in the experiment.}
\label{Tab:tickers}
\begin{tabular*}{0.8\textwidth}{@{\extracolsep{\fill}}ccccccccc}
\toprule
2802 & 3382 & 4063 & 4452 & 4568 & 4578 & 6501 & 6502 & 7203 \\
7267 & 8001 & 8035 & 8058 & 8306 & 8411 & 9202 & 9613 & 9984 \\
\bottomrule
\end{tabular*}
\end{table*}

\subsection{Evaluation Metrics}\label{Sec:evaluation_metrics}
\subsubsection{All-Time High Anomaly}
To evaluate whether our simulation reproduces anomalies observed in real financial markets, we focus on the so-called all-time high anomaly. This anomaly refers to the empirical finding that the nearness of a stock’s current price to its historical maximum tends to negatively predict future returns. When prices approach their all-time highs, subsequent returns are systematically lower---a pattern inconsistent with neoclassical models, which assume that prices reflect only current fundamentals and not past trajectories.

This anomaly exemplifies path dependence, a phenomenon in which historical price dynamics---rather than only present market conditions---affect price formation. Path dependence is thought to arise from psychological mechanisms such as context-dependent loss aversion, where investors evaluate outcomes relative to prior benchmarks (e.g., purchase prices or historical highs).

To assess whether our proposed agent model contributes to the emergence of such path-dependent dynamics, we calculate the ordinary least squares (OLS) estimates of the coefficient $\beta^h$ for each simulation result:
\begin{eqnarray}
\frac{p_{t+T}}{p_t}=\mathrm{const}+\beta^h\frac{p_t}{p_{1:t}^h}
\end{eqnarray}
Li and Yu~\cite{reference_point_ath1} show that $\beta^h < 0$ in empirical stock market data, supporting the presence of the all-time high anomaly. In our evaluation, we estimate $\beta^h$ from both real and simulated data, and compare the results across three return horizons: 10-day, 15-day, and 30-day intervals. We hypothesize that the inclusion of FCLAgents---designed to exhibit psychologically motivated trading behavior through LLM-derived loss aversion---enables the simulation to replicate this anomaly, thereby demonstrating the model’s ability to generate realistic, path-dependent market dynamics.


\subsubsection{Other Stylized Facts}\label{Sec:stylized_facts}
To ensure the validity of our simulations, we assess whether the simulation results exhibit key stylized facts observed in real financial markets. Following \cite{stylized_facts,realism}, we consider the following stylized facts:
\begin{itemize}
\item The kurtosis of the stock return distribution, denoted as $\kappa_r$ is positive.
\item The autocorrelation of the absolute return series, denoted as $\gamma(T)$, remains positive over a wide range of time lags $T$.
\item The correlation between absolute return and execution volume, denoted as $\rho$ is positive.
\end{itemize}
To calculate the above statistics, we resample execution events in the synthetic tick data to match the intraday transaction path observed in real data, enabling the construction of one-minute bar price series on a common time basis.

\section{Results and Discussion}
This section presents the results of the multi-agent trading simulations. We first compare the macro-level statistics introduced in Section~\ref{Sec:evaluation_metrics} to examine the distinct effects of incorporating FCLAgents. We then analyze the action histories of FCLAgents to investigate the behavioral mechanisms underlying the simulation outcomes and to assess the plausibility of their decision-making patterns.


\subsection{Multi-Agent Simulation}
Table~\ref{Tab:sim_results} shows the OLS coefficient of nearness to the all-time high $\beta^h$ calculated on real and simulated data. As the number of FCLAgents $n_{\mathrm{FCL}}$ increases, the estimated $\beta^h$ from the simulations tend to fall within the range observed in the real data. These results indicate that the inclusion of FCLAgents enables the simulation to reproduce the anomaly---previously unreplicable by FCNAgents alone---highlighting the effect introduced by LLM-guided trading intentions. 

Table~\ref{Tab:stylized_facts} summarizes the descriptive statistics calculated for real data and the simulation results. All of the stylized facts mentioned in Sec~\ref{Sec:stylized_facts} were observed in the real data. In the simulation results, the stylized facts were observed regardless of the value of $n_{\mathrm{FCL}}$. This indicates that the introduction of FCLAgents does not compromise the validity of the simulation as a realistic market model.

\begin{table}[tbp]
\centering
\caption{OLS estimates of $\beta^h$ under different simulation conditions and horizons. Each value is scaled by $10^{-1}$. Real data entries include 1-standard deviation intervals across different tickers listed in Table~\ref{Tab:tickers}. Simulation data entries include sample standard deviation across trials.}
\label{Tab:sim_results}
\begin{tabular}{>{\raggedright}p{3.8cm}>{\centering\arraybackslash}p{2.4cm}
                                    >{\centering\arraybackslash}p{2.4cm}
                                    >{\centering\arraybackslash}p{2.4cm}}
\toprule
\textbf{Setting} & \textbf{10-day} & \textbf{15-day} & \textbf{30-day} \\
\midrule
Real Data & \makecell{$-0.73$\\$[-1.27,\,-0.20]$} & \makecell{$-1.05$\\$[-1.78,\,-0.25]$} & \makecell{$-1.49$\\$[-2.72,\,-0.25]$} \\
\addlinespace
Simulation $n_{\mathrm{FCL}} = 0$ & $-0.01~(\pm0.76)$ & $-0.02~(\pm0.78)$ & $-0.03~(\pm0.81)$ \\
Simulation $n_{\mathrm{FCL}} = 1$ & $-0.67~(\pm0.78)$ & $-0.68~(\pm0.84)$ & $-0.91~(\pm1.40)$ \\
Simulation $n_{\mathrm{FCL}} = 2$ & $-0.42~(\pm0.51)$ & $-0.49~(\pm0.67)$ & $-0.79~(\pm1.24)$ \\
Simulation $n_{\mathrm{FCL}} = 3$ & $-0.32~(\pm0.52)$ & $-0.35~(\pm0.52)$ & $-0.62~(\pm1.38)$ \\
Simulation $n_{\mathrm{FCL}} = 4$ & $-1.20~(\pm1.15)$ & $-1.30~(\pm1.46)$ & $-1.69~(\pm1.72)$ \\
Simulation $n_{\mathrm{FCL}} = 5$ & $-1.31~(\pm1.90)$ & $-1.34~(\pm1.90)$ & $-1.64~(\pm2.26)$ \\
\bottomrule
\end{tabular}
\end{table}

\begin{table}[tbp]
\centering
\caption{Stylized facts in real and simulated results. Real data entries include 1-standard deviation intervals. The check mark indicate that the corresponding stylized fact is satisfied, which is described in Section~\ref{Sec:stylized_facts}.}
\label{Tab:stylized_facts}
\begin{tabular}{>{\raggedright}p{3.1cm}>{\centering\arraybackslash}p{1.7cm}
                                    >{\centering\arraybackslash}p{1.7cm}
                                    >{\centering\arraybackslash}p{1.7cm}
                                    >{\centering\arraybackslash}p{1.7cm}
                                    >{\centering\arraybackslash}p{1.7cm}}
\toprule
\textbf{Setting} & $\kappa_r$ & $\gamma(1)$ & $\gamma(5)$ & $\gamma(10)$ & $\rho$ \\
\midrule
Real Data & \makecell{$7.85$ \\ $(\pm1.07)$} & \makecell{$0.19$ \\ $(\pm0.02)$} & \makecell{$0.14$ \\ $(\pm0.02)$} & \makecell{$0.11$ \\ $(\pm0.01)$}  & \makecell{$0.46$ \\ $(\pm0.07)$} \\
\addlinespace
Simulation $n_{\mathrm{FCL}} = 0$ & $5.11$~$\checkmark$ & $0.28$~$\checkmark$ & $0.02$~$\checkmark$ & $0.01$~$\checkmark$ & $0.07$~$\checkmark$ \\
Simulation $n_{\mathrm{FCL}} = 1$ & $5.21$~$\checkmark$ & $0.28$~$\checkmark$ & $0.02$~$\checkmark$ & $0.01$~$\checkmark$ & $0.07$~$\checkmark$ \\
Simulation $n_{\mathrm{FCL}} = 2$ & $5.38$~$\checkmark$ & $0.26$~$\checkmark$ & $0.04$~$\checkmark$ & $0.01$~$\checkmark$ & $0.08$~$\checkmark$ \\
Simulation $n_{\mathrm{FCL}} = 3$ & $6.54$~$\checkmark$ & $0.30$~$\checkmark$ & $0.04$~$\checkmark$ & $0.01$~$\checkmark$ & $0.08$~$\checkmark$ \\
Simulation $n_{\mathrm{FCL}} = 4$ & $8.01$~$\checkmark$ & $0.24$~$\checkmark$ & $0.02$~$\checkmark$ & $0.02$~$\checkmark$ & $0.07$~$\checkmark$ \\
Simulation $n_{\mathrm{FCL}} = 5$ & $5.87$~$\checkmark$ & $0.26$~$\checkmark$ & $0.02$~$\checkmark$ & $0.02$~$\checkmark$ & $0.08$~$\checkmark$ \\
\bottomrule
\end{tabular}
\end{table}

\begin{figure*}[tbp]
\centering
\includegraphics[width=0.6\linewidth]{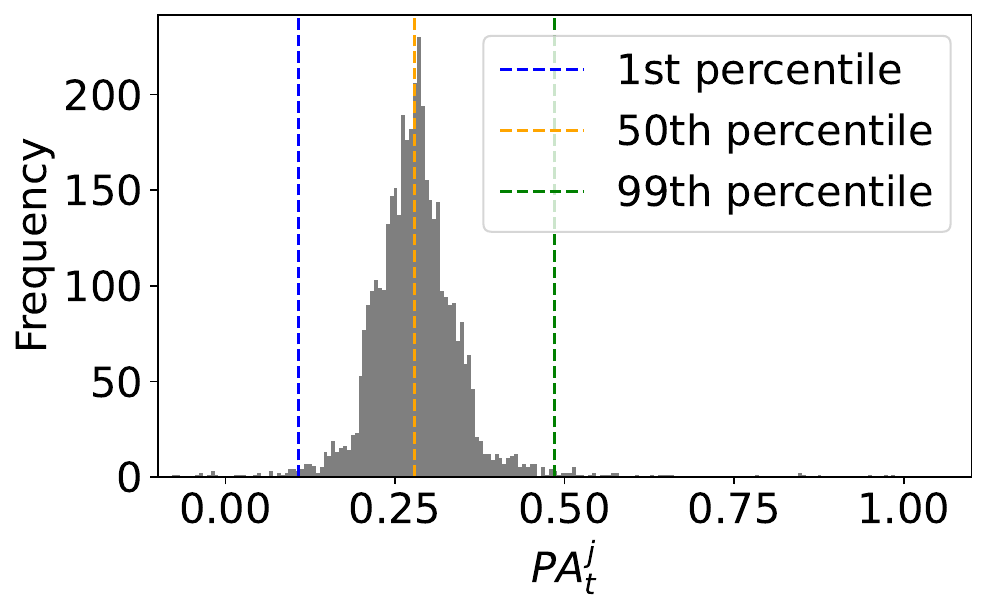}
\caption{Proportion of asset holdings by FCLAgents in a simulation trial ($n_{\mathrm{FCL}}=5$). The sample size is $4,072$. The dotted lines indicate the 1st, 50th, and 99th percentiles of the empirical distribution, corresponding to values of $0.11$, $0.28$, and $0.48$, respectively.}
\label{Fig:pa}
\end{figure*}

Fig.~\ref{Fig:pa} illustrates the histgram of the asset proportion $PA_t^j$, for FCLAgents in a representative trial of the simulation with $n_{\mathrm{FCL}}=5$. 
\begin{align}
PA_t^j=\frac{p_tw_t^j}{c_t^j+p_tw_t^j}
\end{align}
The observed $99\%$ interval of $[0.11,0.48]$ suggests that FCLAgents consistently maintained a balanced portfolio over time, rather than adopting extreme positions such as full cash or full investment. This behavior indicates that the agents’ trading decisions were not arbitrary, but reflected a degree of risk management. These results support the plausibility of FCLAgent as a behaviorally grounded and functionally reasonable component in multi-agent market simulations.

\begin{figure*}[tbp]
\centering
\includegraphics[width=0.6\linewidth]{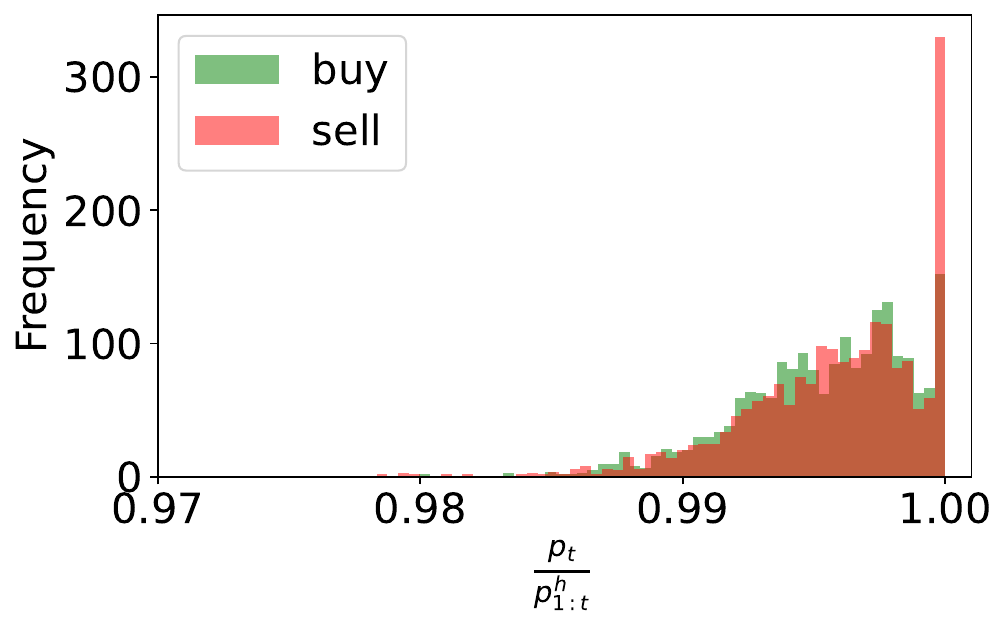}
\caption{All-time high nearness $\frac{p_t}{p_{1:t}^h}$ at the time of the FCLAgents submit orders in a simulation trial ($n_{\mathrm{FCL}}=5$). The sample size are $2,023$ and $2,049$ for purchase and sales, respectively.}
\label{Fig:ath_nearness_fcl_actions}
\end{figure*}

Fig.~\ref{Fig:ath_nearness_fcl_actions} shows the histograms of the all-time high nearness $\frac{p_t}{p_{1:t}^h}$ at the time of stock purchases and sales made by FCLAgents in the same trial of Fig.~\ref{Fig:pa}. The number of purchases ($2,023$) and sales ($2,049$) made by FCLAgents are nearly balanced, indicating no inherent bias toward one action over the other. However, all-time high nearness appears to significantly influence FCLAgents' decision-making, as evidenced by the following observations. First, when the nearness is exactly $1.00$---indicating a new all-time high---there were $102$ purchases compared to $291$ sales, suggesting a strong tendency to sell during price peaks. Second, a Kolmogorov–Smirnov test comparing the nearness distributions of buy and sell decisions yields a p-value of $4.74\times10^{-8}$, and a Mann–Whitney U test yields a p-value of $1.94\times10^{-6}$, both indicating a statistically significant difference in the context under which purchases and sales occur. These results suggest that FCLAgents not only exhibit a minimally reasonable behavioral pattern that ensures the simulation remains consistent with fundamental assumptions of market realism, but also incorporate a specific bias toward all-time high prices. This tendency, in which FCLAgents are more likely to sell near price peaks and buy after declines, may serve as a behavioral mechanism underlying the observed anomaly whereby nearness to the all-time high negatively predicts future returns.

\section{Context-Dpendency of LLMs' Loss Aversion}
In the previous sections, we confirmed that (1)the inclusion of FCLAgents enables the simulation to reproduce the empirical anomaly---the negative correlation between proximity to the all-time high and future returns, and (2) the FCLAgents exhibit reasonable behavioral patterns in terms of balancing portfolio, while the all-time high nearness biases FCLAgents toward selling. Based on these results, we hypothesized that the all-time high anomaly arises from a context-dependent loss aversion. Specifically, we conjectured that when the current price is near the all-time high, LLMs are more likely to realize gains early by selling profitable stocks, while they tend to tolerate additional risk and continue buying when the price has dropped below the peak, which leads to higher subsequent returns in lower nearness regimes.

To further analyze how the all-time high price influence the LLMs' behavioral tendencies, we conducted a single-turn simulation. We asked the LLMs to make a one-time order decision under varying contextual scenarios, while fixing other variables. The following analysis focuses on GPT-4o~\cite{gpt4} and Qwen-2.5 7B~\cite{qwen2} in addition to Llama 3.1 8B.



\begin{figure}[tbp]
\begin{algorithm}[H]
    {\footnotesize
    \caption{{\footnotesize Single-turn simulation for analyzing loss aversion}}
    \label{Alg:single_turn}
    \begin{algorithmic}[1]
        \REQUIRE $t$, the current time, $c_t^j$, the current holding cash amount, $p_1^j$, the initial stock price at which the LLM bought at time $1$, $v_1^j$, initial stock volume which the LLM bought, $r_{min}$ and $r_{max}$, the minimum and maximum stock returns at time $t$, $p_{1:t}^l$ and $p_{1:t}^h$, the all-time high and low prices.
        \STATE Uniformly sample $r_t$ from range $[r_{min},r_{max}]$.
        \STATE $p_t \leftarrow p_1\exp(r_t)$
        \STATE Set current portfolio as $c_t^j$, $w_t^j\leftarrow v_1^j$, and $\bar{g}_t^j\leftarrow v_1(p_t-p_1)$.
        \STATE Set market condition as $p_t$, $p_{1:t}^l$, $p_{1:t}^h$, $T_{sim}-t$, and $T_{sim}$..
        \STATE Set trading history $h_t^j$ as $\{1,p_1^j,v_1^j\}$ and $\forall t'\in\{2,...,t-1\}~ v_{t'}^j=0$.
        \STATE Observe investment decision by the LLM given the current portfolio, market condition, and trading history.
    \end{algorithmic}
    }
\end{algorithm}
\end{figure}

\begin{figure}[tbp]
    \centering
        \includegraphics[scale=.4]{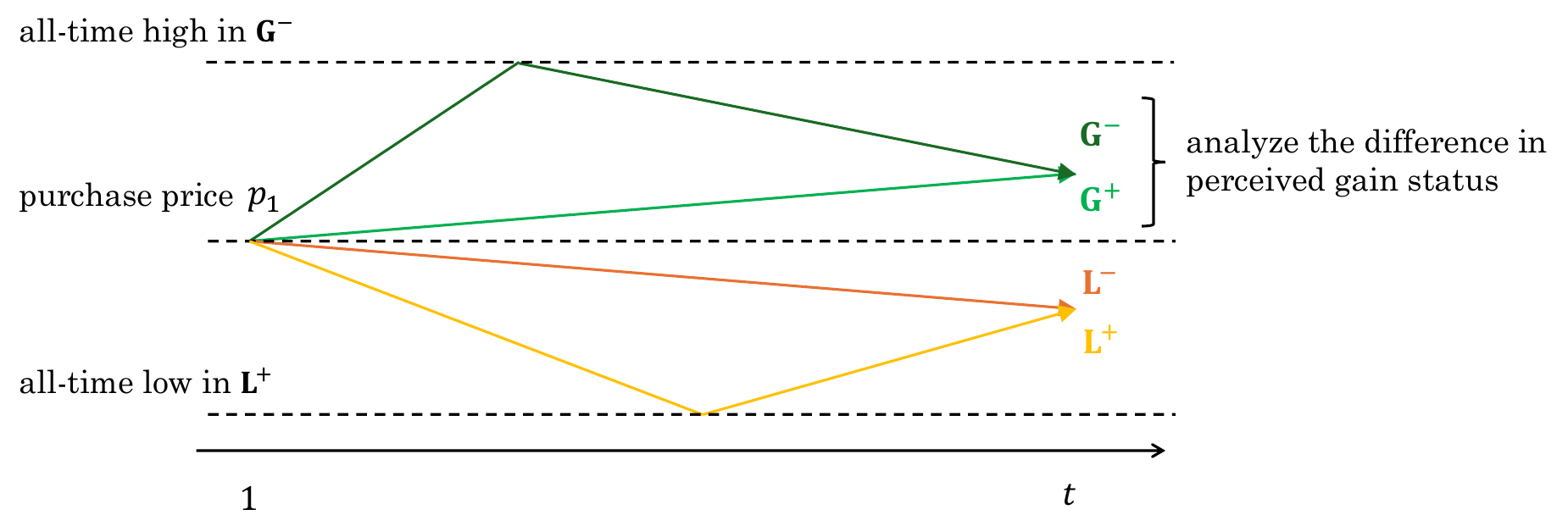}
    \caption{Conceptual diagram illustrating the four situations with different contexts in the single turn simulations. In \textbf{G\textsuperscript{$-$}} and \textbf{L\textsuperscript{$+$}} situation, all-time high (low) price can serve as a reference point, although the current and purchase prices are the same as \textbf{G\textsuperscript{$+$}} and \textbf{L\textsuperscript{$-$}}, respectively.}
    \label{Fig:situations_single_turn}
\end{figure}

In the single-turn simulations, an LLM was asked at time $t$ to make an order decision following the procedure described in Algorithm~\ref{Alg:single_turn}. We set $p_1 = 300.0$, $c_t^j = 30,000$, $v_1^j = 10$, $t = 50$, and $T_{sim} = 100$, and conducted the experiment under four different situations designed to reflect contextual factors influencing loss aversion, as outlined below. Each situation was tested across 100 trials, with a different seed value in each trial.
\begin{itemize}
\item Unrealized \textbf{g}ain and updating all-time high situation (\textbf{G\textsuperscript{$+$}}). $r_{min}=0.0$, $r_{max}=0.5$, $p_{1:t}^h=p_t$.
\item Unrealized \textbf{g}ain and declining from all-time high situation (\textbf{G\textsuperscript{$-$}}). $r_{min}=0.0$, $r_{max}=0.5$, $p_{1:t}^h=p_1\exp(2r_t)$.
\item Unrealized \textbf{l}oss and updating all-time low situation (\textbf{L\textsuperscript{$-$}}). $r_{min}=-0.5$, $r_{max}=0.0$, $p_{1:t}^l=p_t$.
\item Unrealized \textbf{l}oss and recovering from all-time low situation (\textbf{L\textsuperscript{$+$}}). $r_{min}=-0.5$, $r_{max}=0.0$, $p_{1:t}^l=p_1\exp(2r_t)$.
\end{itemize}
Fig.~\ref{Fig:situations_single_turn} illustrates each situation. The situations are categorized as \textbf{G\textsuperscript{$+$}}, \textbf{G\textsuperscript{$-$}}, \textbf{L\textsuperscript{$-$}}, and \textbf{L\textsuperscript{$+$}} to examine how LLMs recognize two types of reference point, the purchase price~\cite{reference_point_bought_price} and the all-time high and low prices~\cite{reference_point_ath2,reference_point_ath1}. For example, in the \textbf{G\textsuperscript{$+$}} situation, the purchase price $p_1^j$ serves as the only reference point. In contrast, in the \textbf{G\textsuperscript{$-$}} situation, all-time high price can also act as a reference point, as discussed in Section~\ref{Sec:loss_aversion_agent}.

\begin{table*}[tbp]
\centering
\caption{Number of buy and sell orders in each situation (Gains or Losses) during the single-turn experiment. Each cell shows: Net Order Flow (Buy, Sell).}
\label{Tab:single_turn_result}
\begin{tabular}{l|>{\centering\arraybackslash}p{2.3cm}
                  >{\centering\arraybackslash}p{2.3cm}
                  >{\centering\arraybackslash}p{2.3cm}
                  >{\centering\arraybackslash}p{2.3cm}}
\toprule
\textbf{LLM Type} & \textbf{G\textsuperscript{$+$}} & \textbf{G\textsuperscript{--}} & \textbf{L\textsuperscript{--}} & \textbf{L\textsuperscript{$+$}} \\
\midrule
GPT-4o            & $-94$ (3, 97)   & $-70$ (15, 85)  & $+80$ (90, 10)  & $+74$ (87, 13) \\
Llama 3.1 8B      & $-80$ (10, 90)  & $-68$ (16, 84)  & $+48$ (74, 26)  & $+30$ (65, 35) \\
Qwen-2.5 7B       & $-72$ (14, 86)  & $-74$ (13, 87)  & $+86$ (93, 7)   & $+60$ (80, 20) \\
\bottomrule
\end{tabular}
\end{table*}

Table~\ref{Tab:single_turn_result} shows the number of buy and sell orders in the \textbf{G\textsuperscript{$+$}}, \textbf{G\textsuperscript{$-$}}, \textbf{L\textsuperscript{$-$}}, and \textbf{L\textsuperscript{$+$}} situations. The results from the situations \textbf{G\textsuperscript{$+$}} and \textbf{G\textsuperscript{$-$}} indicate that the number of sell orders exceeded buy orders, suggesting that the LLMs tended to hastily realize their gains when the stock price rose above their purchase price. In \textbf{G\textsuperscript{$-$}} of GPT-4o and Llama 3.1 8B, compared to \textbf{G\textsuperscript{$+$}}, the number of sell orders slightly decreased, while buy orders increased, reflecting loss aversion when the all-time high price was perceived as a reference point. Despite the unrealized gains remaining positive, the fact that the current price had declined from its peak prompted the LLMs to take on additional risk. In contrast, in situations \textbf{L\textsuperscript{$-$}} and \textbf{L\textsuperscript{$+$}}, the number of buy orders exceeded sell orders. This behavior aligns with loss aversion, where investors are reluctant to realize their losses and continue holding depreciating stocks. Furthermore, in the \textbf{L\textsuperscript{$+$}} situation, where the stock price was recovering from its all-time low, the all types of LLMs tended to increase their sell orders. Similar to \textbf{G\textsuperscript{$-$}}, the all-time low price acted as an additional reference point alongside with the purchase price. However, in the gain scenarios, Qwen-2.5 7B exhibited different tendency to the other LLMs. Qwen-2.5 7B maintained a consistent preference for realizing gains, even after the price declined from its peak. This could reflect a more rigid interpretation of gain status based primarily on the purchase price, rather than incorporating multiple reference points. These results suggest that while some LLMs exhibit context-dependent loss-averse behavior and effectively replicate complex behavioral patterns, others do not. This underscores the importance of verifying, on an individual basis, whether the intended behavioral tendencies are actually reflected in the LLMs’ decision-making, in addition to analyzing the overall simulation outcomes. Without such validation, simulations may appear plausible at the aggregate level while failing to capture the underlying psychological mechanisms they are intended to model.

\section{Conclusion}
We proposed the FCLAgent, an LLM-based agent model designed to serve as a more sophisticated representation of human investors in multi-agent financial market simulations. The FCLAgent integrates context-dependent, human-like behavioral biases---elicited from LLMs---into the agent’s agents’ buy/sell decisions, while relying on rule-based mechanisms to determine order price and volume, thereby circumventing LLMs’ limitations in numerical reasoning. Our simulation experiments demonstrated that incorporating FCLAgents enables the reproduction of empirically observed market anomalies---such as the negative correlation between proximity to an asset's all-time high and its subsequent returns---that traditional agents alone could not replicate. The inclusion of FCLAgents preserved key stylized facts typically used to assess market realism, indicating that LLMs can enrich agent behavior without compromising the overall integrity of the simulation environment. We also conducted additional simulation experiments focusing on the characteristics of LLM-based agent that were not represented in traditional statistical agent models. Our analysis revealed that the loss-averse tendencies of certain types of LLMs are highly context-dependent, shaped by multiple reference points such as purchase price and all-time high or low.

Through this study, we argue that the integration of LLMs into agent-based simulations offers a key advantage: the ability to represent context-dependent behavioral biases that are difficult to capture through conventional approaches. As future work, we aim to enhance the heterogeneity of LLM-based agents by incorporating demographic factors and other contextual inputs. While human investors tend to exhibit systematic behavioral biases such as loss aversion, their concrete responses can vary widely---for example, some may respond to losses by panic-selling, while others may take on additional risk in an attempt to recover. Such behavioral diversity is shaped by a variety of factors, including situational framing, individual profiles, and extensive communications with others. By constructing agent models that more faithfully capture these psychologically rich and context-dependent variations, we enable simulations that can reproduce complex phenomena such as speculative bubbles and market cascades.

\begin{appendix}
\section{Prompt}\label{Appx:prompt}
This section provides examples of the text prompts used in the experiments. Our prompt consisted of premise, instructions, and answer format. Here's an example of the prompts used in our experiments. 

{\footnotesize
\begin{verbatim}
(Premise) You are a participant of the simulation of stock markets. Behave
as an investor. Answer your order decision after analysing the given
information.
(Instruction) Your current portfolio is provided as a following format.
Unrealized gain refers to the increase in value of the investment that has
not yet been sold. It represents the potential profit on your stock
position. Negative unrealized gain means  that the investment has decreased
in value.
[Your portfolio]cash: {}
[Your portfolio]market id: {}, volume: {}, unrealized gain: {}, ...
Each market condition is provided as a following format.
[Market condition]market id: {}, current market price: {},
all time high price: {}, all time low price: {}, ...
[Market condition]market id: {}, remaining time: {}, total time: {}
Your trading history is provided as a following format. Negative volume
means that you sold the stock.
[Your trading history]market id: {}, price: {}, volume: {}, ...
Order flow imbalance is provided as a following format. Order flow imbalance
means the difference between the number of buy and sell orders submitted to
the stock market. Order flow imbalance is calculated as the difference between
the number of buy and sell orders. Order flow imbalance can range from -1 to 1.
Negative order flow imbalance indicates that the number of sell orders exceed
that of buy orders. If the order flow is positive (negative), the fundamental
value tends to be high (low). Higher absolute value of order flow imbalance
indicates that orders are imbalance to one side, and suggests stronger evidence
about the fundamentals value of the stock.
[Order flow imbalance]market id: {}, order flow imbalance: {}, ...
(Information) Here's the information.
[Your portfolio]cash: 30000
[Your portfolio]market id: 0, volume: 10, unrealized gain: -63.0
[Market condition]market id: 0, current market price: 293.7,
all time high price: 300.0, all time low price: 287.5
[Market condition]market id: 0, remaining time: 70, total time: 100
[Your trading history]market id: 0, price: 300.0, volume: 10
[Order flow imbalance]market id: 0, order flow imbalance: 0.01
(Answer format) Decide your investment in the following JSON format. Do not
deviate from the format, and do not add any additional words to your response
outside of the format. Make sure to enclose each property in double quotes.
Order volume means the number of units you want to trade the stock. Possible
is_buy means whether to buy or sell the stock. is_buy must be True or False.
Short selling is not allowed. If your holding stock volume in the portfolio is
negative, buy them back immediately.  Cash shortage is not allowed. If your
cash amount in the portfolio is negative, sell your holding stocks immediately.
Try to keep your order volume as non-zero and not-extreme as possible. Try to
keep your portfolio balanced. If you feel you are holding a lot of stocks or
your cash amount is insufficient, you should sell them. Order price means the
limit price at which you want to buy or sell the stock. By adjusting order
price, you can trade at a more favorable price or adjust the time it takes to
execute a trade. Here are the answer format.
{"<market id>": {"order_price": "<order price>", "is_buy": "<True or False>",
"order_volume": "<order volume>", "reason": "<reason>"} ...}
Now, decide your order. Please explain the reason and your emotion in
as much detail as possible.
\end{verbatim}}

When $c_t^j<0$ or $w_t^j<0$, we added the warning to the agent in the prompt:

{\footnotesize
\begin{verbatim}
[Your portfolio]cash: -1000 (Caution! Your cash amount is negative!
To avoid this situation, you have to sell the stocks.)
\end{verbatim}}

{\footnotesize
\begin{verbatim}
[Your portfolio]market id: 0, volume: -1, unrealized gain: 3.0 
(Caution! Your holding stock volume is negative! To avoid this
situation, you have to buy this stock.)
\end{verbatim}}

\end{appendix}

\begin{credits}
\subsubsection{\ackname}
This work was supported by JSPS KAKENHI Grant Number JP25KJ1124.
\end{credits}

{\footnotesize
\bibliographystyle{splncs04}
\bibliography{citation}
}

\end{document}